\documentclass[aps,pre,amsmath,amssymb,reprint,superscriptaddress,floatfix]{revtex4-2}

\usepackage[utf8]{inputenc}
\usepackage{graphicx}
\usepackage{dcolumn}
\usepackage{bm}
\usepackage[hidelinks]{hyperref}
\usepackage{newtxtext,newtxmath}

\begin{document}

\title{Steering toward $\mathbb{Z}_2$-symmetric states in monitored Clifford circuits with two-qubit feedback} 

\author{Yunsu Kim}
\affiliation{Department of Physics and Photon Science, Gwangju Institute of Science and Technology, Jeonnam-Gwangju 61005, Republic of Korea}

\author{Su-Chan Park}
\email{spark0@catholic.ac.kr}
\affiliation{Department of Physics, The Catholic University of Korea, Bucheon 14662, Republic of Korea}

\author{Dong-Hee Kim}
\email{dongheekim@gist.ac.kr}
\affiliation{Department of Physics and Photon Science, Gwangju Institute of Science and Technology, Jeonnam-Gwangju 61005, Republic of Korea}

\date{\today}

\begin{abstract}
Intermittent measurements and feedback controls on random quantum circuits can provide an alternative platform for studying nonequilibrium phase transitions. We present two-qubit gate controls conditioned on measurement outcomes, combining feedback toward two symmetric absorbing states with probabilistic finite-range reset of feedback flags to facilitate scrambling. By demonstrating this protocol on random Clifford circuits, we find that the system undergoes an entanglement transition at a lower critical measurement rate when the finite-range reset is turned on, followed by an absorbing-state transition at a higher critical measurement rate. We characterize critical behavior along the phase boundaries using finite-size scaling analysis with various initial states, showing that critical exponents at the absorbing-state transition are consistent with those in the directed Ising universality class of parity-conserved branching-annihilating random walks in one dimension.
\end{abstract}

\maketitle

\section{\label{sec:intro}Introduction}
Random quantum circuits~\cite{Nahum2017, Nahum2018, Potter2022, Fisher2023} can serve as a highly controllable platform for studying nonequilibrium phase transitions because of their flexibility in realizing various dynamical conditions. The measurement-induced phase transition (MIPT)
\cite{Skinner2019, Li2019, Gullans2020a, Gullans2020b}
is a prime example, which arises from the competition between quantum scrambling due to random unitary gates and entanglement destruction induced by interspersed local measurements on qubits.  
Characterizing these dynamical phase transitions between volume-law and area-law entangled states in monitored random circuits has attracted a great deal of attention,
both theoretical and experimental
\cite{Li2018, Chan2019, Skinner2019, Li2019, Zabalo2020, Bao2020, Jian2020, Choi2020, Turkeshi2020, Fuji2020, Gullans2020a, Gullans2020b, Iaconis2020, Lavasani2021a, Ippoliti2021, Nahum2021, Gopalakrishnan2021, Sang2021, Lunt2021, Buchhold2021, Lavasani2021b, Minato2022, Block2022, Muller2022, Zabalo2022, Han2022, Sierant2022, Noel2022, Sierant2023, Koh2023, Piroli2023, Zabalo2023, Ravindranath2023, Shkolnik2023, Friedman2023, ODea2024, Tikhanovskaya2024, Ha2024, Kamakari2025, Pokharel2026, Wu2026}.

Beyond random unitary dynamics, monitored circuits combined with feedback have emerged as promising candidates for quantum state steering~\cite{Grover2000, Shende2006, Plesch2011, Zhang2021, Zhang2022}, a basic prerequisite for implementing quantum error correction~\cite{Gottesman1997, Nielsen2010, Roffe2019, Putterman2025}, various quantum algorithms, and Hamiltonian simulations~\cite{Berry2007, Berry2015, Low2017, Low2019}. Feedback controls are conditional gate operations based on midcircuit readouts~\cite{Sierant2023, Ravindranath2023, Piroli2023, Friedman2023, ODea2024, Makki2024, LeMaire2024, Ravindranath2025, Ha2026}, allowing the precise steering of a quantum state toward a desired target. If these feedback mechanisms are strictly designed to drive the system into a target state and prevent it from escaping, the resulting dynamics exhibit phenomena closely analogous to absorbing-state phase transitions (APTs) found in classical systems~\cite{Hinrichsen2000, Odor2004, Sieberer2025}. Furthermore, because such feedback-assisted state steering offers a promising avenue to circumvent the post-selection problem—one of the primary obstacles to experimentally observing MIPTs—these adaptive protocols have garnered extensive theoretical interest~\cite{Ippoliti2021, Sierant2023, ODea2024, Makki2024} alongside rapid experimental progress toward their physical realization~\cite{Sierant2022, Kamakari2025, Wu2026, Pokharel2026}.

Because the accessibility of absorbing states is likewise governed by the competition between opposing dynamical processes, the presence of symmetries has been one of the main ingredients in APT studies, as it can fundamentally alter the universality class of the transition~\cite{Hinrichsen2000, Odor2004}. A paradigmatic example is provided by one-dimensional branching-annihilating random walks (BARWs)~\cite{Grassberger1984, Takayasu1992, Jensen1993a, Jensen1993b, Jensen1994, Cardy1996}. When particles branch into an odd number of offspring (BARW-o), spontaneous annihilation effectively emerges through a combination of branching and pairwise annihilation even if it is not explicitly included in the microscopic rules, placing the transition in the directed percolation (DP) universality class~\cite{Takayasu1992, Jensen1993a}. In contrast, when branching produces an even number of offspring (BARW-e), particle-number parity is conserved and spontaneous annihilation is prohibited; consequently, a system initialized with an odd number of particles can never reach the empty absorbing state regardless of the branching, annihilation, or diffusion rates, driving the transition into a distinct universality class. Although this universality class is variously referred to in the literature as directed percolation 2 (DP2), parity-conserving, or generalized voter class, we refer to it throughout this paper as the directed Ising (DI) universality class~\cite{Grassberger1984, Takayasu1992, Jensen1993b, Jensen1994, Cardy1996, Hwang1998, Menyhard2000, Park2013, Park2020b, Odea2026, Grassberger2026}.

Motivated by this sensitivity of classical APTs to underlying symmetries, symmetry-constrained monitored quantum circuits have been actively explored across various settings. For example, $\mathbb{Z}_2$-symmetric quantum automaton circuits built from Clifford gates exhibit an entanglement transition whose dynamical exponent $z$ matches that of the DI class~\cite{Han2022}, though without aiming at quantum state steering. Monitored circuits with onsite $\mathbb{Z}_2$ symmetry—built from $\hat{Z}_i\hat{Z}_{i+1}$ and $\hat{X}_i$ measurements, and further extended with $\mathbb{Z}_2$-symmetric Clifford gates—have also been studied using linear cross-entropy to probe transitions among spin-glass, paramagnetic, and volume-law phases~\cite{Tikhanovskaya2024}, whereas imposing $\mathrm{U}(1)$ charge conservation on Clifford gates suppresses the volume-law phase and eliminates the MIPT~\cite{Ha2024}, illustrating how symmetry constraints affect entanglement growth.

Within feedback-induced steering, protocols restricted to two-site $\hat{X}_i\hat{X}_{i+1}$ and single-site $\hat{X}_i$ operations yield a DI-class APT in a qubit system and a distinct APT in a qutrit system~\cite{Ha2026}, although they were not designed to support an MIPT. Haar-random circuits combining three-qubit unitary scrambling with two-site projective $Z$ measurements and conditional Pauli-$X$ feedback display both an MIPT and a DI-class APT of domain walls~\cite{Ravindranath2023}. However, replacing the three-qubit gates with two-qubit gates drives the system into the absorbing state at any nonzero measurement rate, thereby eliminating the APT. Moreover, because simulating highly entangled states in Haar-random circuits is computationally demanding at large system sizes, precise determination of the MIPT critical point and the long-time behavior of the volume-law phase remained limited.

In this work, we investigate the adaptive quantum dynamics that drives a qubit chain from an unpolarized state toward spin-polarized states (specifically, two $\mathbb{Z}_2$-symmetric absorbing states, $|\uparrow\uparrow\dots\uparrow\rangle$ or $|\downarrow\downarrow\dots\downarrow\rangle$). To ensure numerical tractability when studying the entanglement structure at large system sizes, we employ random Clifford circuits~\cite{Gottesman1997, Gottesman1998, Nielsen2010, Gottesman2004} consisting of layers of probabilistic projective measurements and conditional feedback-control operations aimed at aligning adjacent spins. By introducing auxiliary classical labels (``flags''~\cite{Sierant2023}) to keep track of local alignment history, we incorporate random two-qubit Clifford gates into the feedback steps without resorting to three-qubit operations. Tuning the measurement rate in this setup induces both the APT toward the $\mathbb{Z}_2$-symmetric states and the MIPT, enabling a scalable finite-size analysis of their critical properties.

Without keeping track of the recent feedback history, the volume-law phase completely vanishes at any nonzero measurement rate $p$ despite the presence of entangling Clifford gates, implying the absence of the MIPT. Instead, the dynamics is governed solely by the APT of domain walls. By evaluating order parameters across three distinct initial states, we confirm that this transition belongs to the 1D DI class. However, upon introducing a nonzero ``unflagging'' rate $u$, which reflects the breaking of local alignment after the application of a random Clifford gate, both the MIPT and the APT emerge sequentially as $p$ increases. By constructing the ($p, u$) phase diagram, we identify critical boundaries for both transitions, delineating the sufficient conditions for the APT and the necessary conditions for the MIPT within this adaptive quantum dynamics framework. Furthermore, while the APT robustly belongs to the 1D DI class regardless of the unflagging rate $u$, tracing the phase boundary—where the critical measurement rate varies with $u$—suggests the presence of a nontrivial crossover occurring within the DI class.

\section{Model and Methods} 
\label{sec:methods}

\subsection{Hybrid Clifford circuit with two classical flags} 
\label{subsec:circuit} 

We consider a $(1+1)$-dimensional brick-wall circuit of random two-qubit Clifford gates.
Each qubit $i$ is augmented with a classical measurement flag $f_i \in \{0,1\}$ that records the outcome of the most recent $Z$ measurement (or the initial configuration), where $f_i=0\,(1)$ corresponds to $|\uparrow\rangle\,(|\downarrow\rangle)$.
From the set $\{f_i\}$ we define a nearest-neighbor flag $g_i$ on the virtual bond $(i,i+1)$, 
which is updated after every circuit layer as $g_i = \neg\bigl(f_i \oplus f_{i+1}\bigr)$.
We refer to $g_i$ as a ``domain-wall'' flag: $g_i=1$ when adjacent flags agree and $g_i=0$ when they differ, signaling a domain wall between sites $i$ and $i{+}1$.
Note that, in this work, we focus on the domain-wall density defined from the quantum state of the qubits rather than directly from the classical flags; nevertheless, we have verified that the classical flag $\{g_i\}$ displays the same critical behavior at the absorbing phase transition.


\begin{figure}
\includegraphics[width=0.48\textwidth]{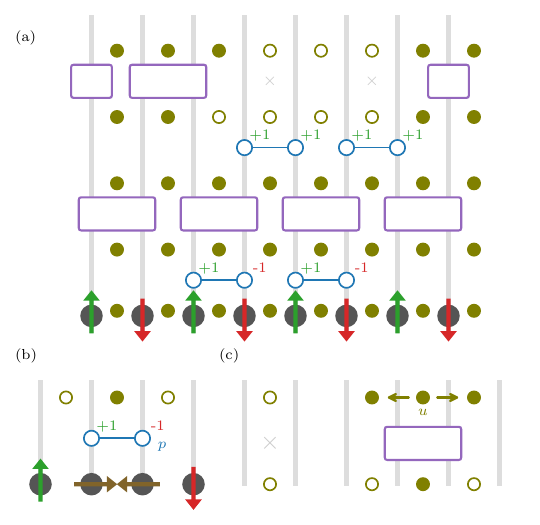}%
\caption{\label{fig:Double-flag_circuit} Double-flag circuit. (a) Schematic of a single time step of the double-flag circuit acting on the fully-occupied initial condition under the periodic boundary condition ($L=8$). (b) Local measurements occur with a probability $p$, after which the domain-wall flags $\{g_i\}$ are updated based on the measurement flags $\{f_i\}$. Each flag $g_i$ indicates whether the adjacent spins are aligned ($g_i=1$, open circle) or not ($g_i=0$, filled olive circle). (c) Feedback-control operations are conditioned on the flag $g_i$: if $g_i = 1$, the qubit pair is left untouched. Otherwise ($g_i = 0$), a random two-qubit Clifford gate is applied to that pair of qubits, and adjacent flags are subsequently reset to $g_{i-1} = g_{i+1} = 0$ with a probability $u$.}
\end{figure}

In the brick-wall structure, one unit-time step consists of two consecutive 
layers of the unitaries. Our protocol, which intersperses measurements
with feedback, proceeds as follows. At each layer, the projective measurement 
of $\hat{Z}_i\hat{Z}_{i+1}$ is performed on the corresponding pair of qubits with probability $p$,
immediately before the two-qubit unitary is applied, as illustrated 
in Fig.~\ref{fig:Double-flag_circuit}(a).
After one sweep of probabilistic measurements, the flags $\{f_i\}$ and $\{g_i\}$ 
are updated according to the measurement outcomes. The subsequent feedback operation
is conditioned on $\{g_i\}$. For the bricks associated with $g_i = 1$, we replace such bricks 
with the identities; otherwise, we keep random Clifford unitaries.
This dynamics drives the system toward one of the two absorbing states, 
$|\uparrow \uparrow \cdots \uparrow \rangle$ or
$|\downarrow \downarrow \cdots \downarrow \rangle$, when $p$ is above a critical value 
$p_c$ as we demonstrate below. 


Crucially, since the application of a random Clifford gate that scrambles the local state likely breaks the alignment with neighboring unmeasured qubits, the circuit must reflect this update in its feedback history. Therefore, after applying the random unitary gate, the adjacent domain-wall flags are ``unflagged'' by setting $g_{i-1} = g_{i+1} = 0$ with an unflagging rate $u\in[0,1]$. In this way, the double-flag circuit seamlessly incorporates the spatial propagation of the feedback history. The third and fourth layers repeat this sequence for even-odd links. Consequently, introducing these flags allows us to include the entangling Clifford gates in the feedback step, even though they may not preserve the target states ($|\uparrow\uparrow\rangle$ or $|\downarrow\downarrow\rangle$) by themselves. In addition, the parity in the number of $0$'s in the domain-wall flags $\{g_i\}$ is not conserved during the circuit evolution. While this might seemingly imply a breakdown of parity conservation for domain walls, $\{g_i\}$ acts just as a proxy for locating them. Ultimately, the two symmetric absorbing states separated by an infinite barrier remain strictly unaffected, regardless of unflagging operations.

Recognizing that the double-flag circuits experience an interplay between quantum scrambling by feedback and wavefunction collapse via projective measurements, it is natural to expect them to exhibit the MIPT at some critical measurement rate. Concurrent with this entanglement phase transition, the specific design of our feedback-control operations drives the system toward $\mathbb{Z}_2$-symmetric absorbing states, thereby inducing the APT of domain walls at another critical measurement rate. Hereafter, we use the subscripts $c_1$ and $c_2$ to denote the critical points of the MIPT and the APT, respectively. This notation reflects the fact that the MIPT always precedes the APT as the measurement rate $p$ increases at a fixed unflagging rate $u$.

\subsection{Initial states} \label{subsec:initial}

As this adaptive quantum dynamics is fundamentally a nonequilibrium process, the choice of initial states and boundary conditions is a critical determinant for characterizing the APT~\cite{Hinrichsen2000, Odor2004}. Motivated by the fact that the transition is driven by the stochastic evolution of domain walls, we focus on three different initial states based on their domain-wall configurations: (i) the fully-occupied initial condition (FOIC), a state with a maximal domain-wall density (e.g., $|\uparrow\downarrow\uparrow\downarrow\dots\rangle$); (ii) the two-particle initial condition (TPIC), featuring a localized pair of domain walls at the center of the chain (e.g., $|\dots\uparrow\uparrow\downarrow\uparrow\uparrow\dots\rangle$); and (iii) the single-particle initial condition (SPIC), containing a single domain-wall interface at the middle of the chain (e.g., $|\dots\uparrow\uparrow\downarrow\downarrow\dots\rangle$). While periodic boundary conditions (PBCs) are applied to the FOIC and the TPIC, open boundary conditions (OBCs) are employed for the SPIC to accommodate its interfacial geometry. To approximate an infinite-size limit in the TPIC and the SPIC simulations, $L$ is chosen such that domain walls never hit either end of the chain during the observation time. We refer to simulations from the FOIC as static simulations and simulations from either the TPIC or the SPIC as defect simulations~\cite{Park2013}.
 
In the regime of low measurement rates ($p < p_{c_{2}}$), the domain walls tend to persist due to the scrambling effect of random unitaries, resulting in an active phase. Conversely, with a sufficiently high measurement rate ($p > p_{c_{2}}$), local alignment dominates, successfully steering the system into the absorbing states in the long-time limit. While these absorbing states (vacuum states of domain walls) are accessible only from the FOIC and the TPIC, the evolution from the SPIC is restricted to a single domain-wall state even in the absorbing phase.

\subsection{\label{sec:observables}Observables}
To characterize the two distinct phase transitions in the double-flag circuit, it is essential to employ appropriate observables tailored to each phenomenon. First, to quantify the entanglement structure of the circuit and identify the MIPT, we evaluate the von Neumann entanglement entropy $S_A = -\text{Tr}_A [\rho_A \log_2{\rho_A}]$ where $\rho_A=\text{Tr}_{A^c}|\psi(T)\rangle\langle\psi(T)|$ is the reduced density matrix of subsystem $A$. Unless otherwise specified, we choose the subsystem $A$ to be a contiguous half of the system. In random Clifford circuits, $S_A$ is identical to any Rényi entropy of any index $n$. Under proper gauge fixing (the ``Clipped'' gauge~\cite{Nahum2017, Li2019, Gullans2020b, Sang2023}), $S_A$ can be efficiently calculated as follows~\cite{Nahum2017, Li2019, Gullans2020b, Sang2023}:
\begin{equation}
\begin{split}
S_A = \frac{1}{2}\big\vert\big\{\tilde{g}_i : [ l(\tilde{g}_i) \in A \text{ and } r(\tilde{g}_i) \in A^c ] & \\
 \text{or } \; [l(\tilde{g}_i) \in A^c \text{ and } r(\tilde{g}_i) \in A] \big\}\big\vert,
\end{split}
\end{equation}
where $\tilde{g}_i$ is the stabilizer generator, $l(\tilde{g}_i)$ and $r(\tilde{g}_i)$ are the left and right endpoints, defined as the leftmost and rightmost sites where $\tilde{g}_i$ acts nontrivially.

To precisely determine $p_{c_{1}}$, we utilize the tripartite mutual information (TMI)~\cite{Zabalo2020, Bao2020, Sierant2023, Zabalo2023, Shkolnik2023, Sang2023}, defined as $I_3 = S_A+S_B+S_C - S_{A\cup B} - S_{B\cup C} - S_{C\cup A} + S_{A \cup B\cup C}$, where $A, B$, and $C$ are contiguous quadrants of the system under PBC. While the TMI is generally more robust against finite-size effects than $S_A$ or the bipartite mutual information, its probability distribution exhibits a heavy tail in the volume-law phase~\cite{Zabalo2023, Shkolnik2023}. Therefore, instead of the mean value of $I_3$, we adopt the probability that $I_3$ is exactly zero, denoted as $P[I_3 = 0]$. In the deep volume-law phase ($p < p_{c_{1}}$), $P[I_3 = 0]$ vanishes, whereas it approaches unity in the deep area-law phase. Near the critical point, $P[I_3 = 0]$ is expected to obey the following scaling ansatz~\cite{Zabalo2023, Shkolnik2023}:

\begin{equation}
\label{eq:SA_prob_TMI}
    P[I_3 = 0] = \mathcal{F} \left( (p - p_{c_{1}}) L^{1/\nu_{\perp, c_{1}}} \right),
\end{equation}
where $\mathcal{F}$ is a universal scaling function and $\nu_{\perp, c_{1}}$ is the correlation length exponent of the MIPT.

Additionally, the MIPT can also be probed via a purification transition, where the system, starting from a mixed state, purifies at a rate governed by the measurement rate~\cite{Gullans2020a, Gullans2020b}. Motivated by this correspondence, we introduce a reference ancilla qubit that is maximally entangled (forming a Bell pair) with a single qubit in the system. After evolving the system through encoding and decoding steps, we evaluate the average bipartite entanglement entropy of this ancilla qubit $\langle S_Q \rangle$, which serves as an order parameter for the entanglement transition.

Turning our attention to the APT, the primary observables are the domain-wall density for static simulations,
\begin{equation}
    \hat{n}_{\text{dw}} (T) \equiv \frac{1}{L}\sum_{i=1}^{L}\left\langle\psi(T)\left|\frac{1-\hat{Z}_i\hat{Z}_{i+1}}{2}\right|\psi(T)\right\rangle,
\end{equation}
and the number of surviving domain walls for defect simulations,
\begin{equation}
    \hat{N}_{\text{dw}}(T)\equiv \sum_{i=1}^{L-1} \left\langle\psi(T)\left|\frac{1-\hat{Z}_i\hat{Z}_{i+1}}{2}\right|\psi(T)\right\rangle,
\end{equation} where $\hat{Z}_i$ denotes the Pauli-$Z$ operator acting on the $i$th qubit.
In fact, $\hat{n}_{\text{dw}}$ corresponds to the quantum analog of the kink density, which acts as the order parameter in the nonequilibrium kinetic Ising model \cite{Menyhard2000}.

Additionally, we evaluate the mean squared distance between the two most distant domain walls $R^2$ from the SPIC, and the steady domain-wall density $\hat{n}_{\text{s}}\equiv \lim\limits_{T\to\infty}{\hat{n}_{\text{dw}}(T)}$ in the active phase from the FOIC; these are used to determine the dynamical exponent $z$ and the order parameter exponent $\beta$, respectively. In the vicinity of the critical point, the temporal behaviors of these observables obey the scaling ansatz as follows~\cite{Park2013}:
\begin{align}\label{eq:SA_Obs_APT}
    \hat{n}_{\text{dw}}(T)&\sim T^{-\delta} f(T/L^z, \Delta T^{1/\nu_{\parallel}}), \\
    \hat{N}_{\text{dw}}(T)&\sim T^{\eta} g(\Delta T^{1/\nu_{\parallel}}),\\
    R^{2}(T)&\sim T^{2/z}h(\Delta T^{1/\nu_{\parallel}}),
\end{align} where $\Delta\equiv p_{c_{2}} - p$ is the distance from the critical point and $f, g$, and $h$ are the scaling functions. The steady domain-wall density in the active phase will follow the scaling ansatz $\hat{n}_{\text{s}}\sim\Delta^{\beta}$. The order parameter exponent $\beta$ is dictated by the relation between critical exponents $\delta$ and $\nu_{\parallel}$; $\beta = \delta\nu_{\parallel}$ \cite{Hinrichsen2000, Odor2004}. The critical exponents are derived from the effective exponents of the observables:
\begin{align}
    \delta_{\text{eff}}&\equiv \log_{b}{\left(\frac{\hat{n}_{\text{dw}}(T)}{\hat{n}_{\text{dw}}(bT)}\right)},\\
    \eta_{\text{eff}}&\equiv \log_{b}{\left(\frac{\hat{N}_{\text{dw}}(bT)}{\hat{N}_{\text{dw}}(T)}\right)},\\
    z_{\text{eff}}&\equiv\frac{2}{\log_{b}{\left(\frac{R^2(bT)}{R^2(T)}\right)}},
\end{align} where $b$ is a constant larger than $1$. In this paper, we set $b=8$. Our estimates for the critical exponents $\delta, \eta_2, \eta_1$ and $z$ should satisfy the generalized hyperscaling relation \cite{Mendes1994} $\delta + \delta' + \eta = d/z$,  where $\delta'$ is the critical exponent associated with the survival probability of domain walls and $d = 1$ is the spatial dimension. It is worth noting that this scaling relation holds regardless of the parity of the initial number of domain walls. If the initial number of domain walls is odd, at least one domain wall always survives, leading to $\delta' = 0$ and consequently $\eta_1 \approx \delta$. Conversely, if the initial number of domain walls is even, we expect $\eta_2 = 0$ and $\delta' = \delta$ due to the duality relation \cite{Mussawisade1998}.

To ensure high precision, we account for corrections to scaling, which provide more accurate estimates of the asymptotic critical exponents. For example, the effective exponent $\delta_{\text{eff}}$ takes the form:
\begin{equation}
\delta_{\text{eff}}=\delta + \frac{c_{\hat{n}}}{\ln{b}}(b^{\chi_{\hat{n}}} - 1) T^{-\chi_{\hat{n}}}+o(T^{-\chi_{\hat{n}}}),
\end{equation}
where $c_{\hat{n}}$ is a non-universal constant, $\chi_{\hat{n}}$ is the leading correction to the scaling exponent (LCSE) of $\hat{n}_{\text{dw}}$, and $o(x)$ stands for all terms that approach zero faster than $x$ for small $x$. To estimate the value of $\chi_{\hat{n}}$, we examine the leading-correction-to-scaling function \cite{Park2013, Park2020b}:
\begin{align}
    \label{eq:LCSEf}
    \Theta_{\hat{n}}(T)&\equiv\left|\frac{\hat{n}_{\text{dw}}(T)\hat{n}_{\text{dw}}(b^2T)}{\hat{n}_{\text{dw}}(bT)^2}-1\right|\nonumber\\
    &= \left|c_{\hat{n}}\right| (b^{\chi_{\hat{n}}} - 1)^2 T^{-\chi_{\hat{n}}} + o(T^{-\chi_{\hat{n}}}).
\end{align}

For an efficient estimation of the effective exponent and the correction-to-scaling function, the observables are measured at the following time steps $T_i$ during circuit evolution:

\begin{equation}
T_i = 
\begin{cases}
i&\textrm{for}~1\leq i\leq40, \\
\lfloor 40 \times 2^{(i-40)/15} \rfloor&\textrm{for}~41\leq i\leq55, \\
2T_{i-15}&\textrm{for}~i\geq56,
\end{cases}
\end{equation}
where $\lfloor x \rfloor$ is the floor function. This choice of measurement intervals yields roughly 50 points per decade on a logarithmic scale. In addition, adopting $b=2^k$ where $k = 1, 2, \dots$ facilitates a straightforward computation of the effective exponents.

\section{\label{sec:results}Results}

\begin{figure}
\includegraphics[width=0.48\textwidth]{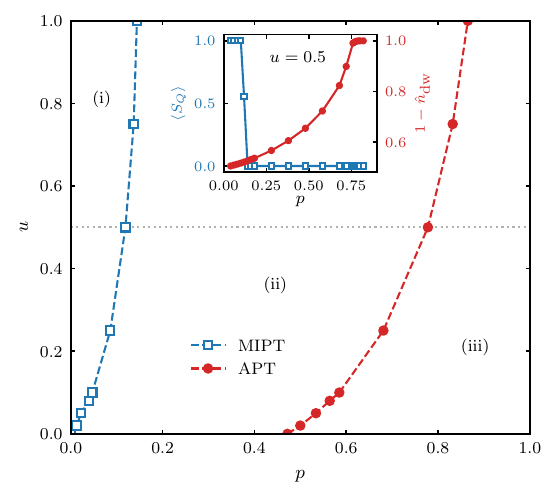}%
\caption{\label{fig:PD} $(p, u)$ phase diagram of double-flag circuits. Three phases are separated by two distinct critical lines: (i) the volume-law and active phase; (ii) the area-law and active phase; and (iii) the area-law and absorbing phase. The inset shows the variation of the two different order parameters, $\langle S_Q\rangle$ and $1-\hat{n}_{\text{dw}}$ for the system size $L=512$ as a function of the measurement rate $p$ when the unflagging rate is fixed at $u = 0.5$ (denoted by the gray dotted line). $\langle S_Q\rangle$ is measured after $L$ encoding and $L$ decoding steps, while $\hat{n}_{\text{dw}}$ is measured at $T=\lfloor{L^{1.75}}\rfloor$.}
\end{figure}

Here we present the numerical results for the double-flag circuits. As summarized in the phase diagram in Fig.~\ref{fig:PD}, the system exhibits three well-defined phases spanned by the measurement rate $p$ and the unflagging rate $u$. When the circuit initialized in the FOIC evolves for a time $T\sim\mathcal{O}(L)$, its bipartite entanglement entropy $S_A$ reaches a saturation value. The critical line of the MIPT (blue dashed line in Fig.~\ref{fig:PD}) distinguishes the volume-law entangled phase, where $S_A \sim \mathcal{O}(L)$, from the area-law entangled phase, where $S_A \sim \mathcal{O}(1)$. At larger measurement rates, the critical line of the APT (red dashed line in Fig.~\ref{fig:PD}) further separates the active phase, characterized by a finite steady domain-wall density $\hat{n}_{\text{s}}$, from the absorbing phase, where the density decays as $\hat{n}_{\text{dw}}\sim T^{-0.5}$. As depicted in the inset of Fig.~\ref{fig:PD}, the occurrence of these two phase transitions can be explicitly confirmed by tracking the respective order parameters as a function of $p$ at a fixed $u > 0$. Unless otherwise specified, all numerical data presented in this section are averaged over at least $1,000$ independent circuit realizations.

\subsection{\label{sec:MIPT}Measurement-induced phase transition}

\begin{figure}
\includegraphics[width=0.48\textwidth]{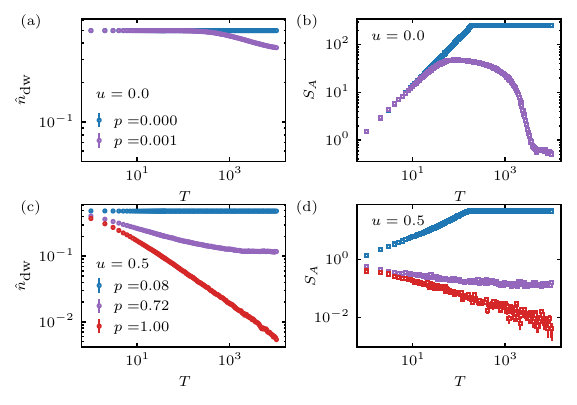}%
\caption{\label{fig:two_OP} Time evolution of the double-flag circuit of length $L=512$ without unflagging operations ($u = 0$) and in the presence of unflagging operations ($u = 0.5$). Log-log plot of the domain-wall density $\hat{n}_{\text{dw}}$ [(a), (c)], and the bipartite entanglement entropy $S_A$ versus $T$ [(b), (d)]. At $u = 0$, a volume-law entangled steady state cannot be achieved even at an infinitesimally small measurement rate $p>0$ due to the eventual decay of $\hat{n}_{\text{dw}}$. At $u = 0.5$, the three colors represent the three phases depicted in Fig.~\ref{fig:PD}: blue for phase (i), purple for phase (ii), and red for phase (iii).}
\end{figure}

Let us first delve into the entanglement phase transition. Figure~\ref{fig:two_OP} illustrates the time evolution of the domain-wall density $\hat{n}_{\text{dw}}$ and the bipartite entanglement entropy $S_A$ for a system size of $L=512$ starting from the FOIC. A distinctive feature is that, in the absence of unflagging operations ($u = 0$), the domain-wall density continuously decays even at infinitesimally small $p$. As a result, the entanglement entropy fails to sustain a volume-law scaling in the steady state, implying that the unflagging mechanism is strictly required for the circuits to exhibit the MIPT.

In contrast, in the presence of unflagging operations, the system exhibits three well-defined phases depending on the measurement rate, as illustrated in Fig.~\ref{fig:two_OP}. For small measurement rates ($p < p_{c_{1}}$), the domain-wall density $\hat{n}_{\text{dw}}$ remains constant around $0.5$, placing the system deep in the active phase. Sustained by this sufficiently high density of domain walls, the steady state of $S_A$ exhibits volume-law entanglement. At intermediate rates ($p_{c_{1}}< p < p_{c_{2}}$), $\hat{n}_{\text{dw}}$ saturates to a smaller but still finite value, keeping the system in the active phase. Nevertheless, the reduced domain-wall density is no longer enough to overcome the effect of measurements, causing the steady-state entanglement entropy to obey an area-law scaling. Finally, for sufficiently large measurement rates ($p > p_{c_{2}}$), the system enters the absorbing phase, where the domain-wall density decays as $\hat{n}_{\text{dw}}\sim T^{-0.5}$, with $S_A$ strictly confined to the area-law regime.

\begin{figure}
\includegraphics[width=0.48\textwidth]{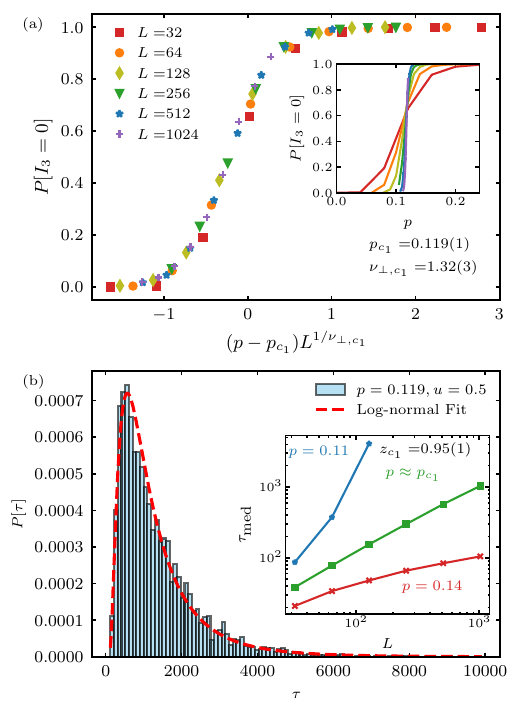}%
\caption{\label{fig:MIPT_0_5} Criticality of the MIPT. (a) The scaling collapse of the probability that the tripartite mutual information becomes exactly zero, $P[I_3 = 0]$, at a fixed unflagging rate $u = 0.5$. According to the scaling ansatz in Eq.~\eqref{eq:SA_prob_TMI}, the scaling collapse yields the critical measurement rate $p_{c_{1}} = 0.119(1)$ and the correlation length exponent $\nu_{\perp, c_{1}}=1.32(3)$. The numbers in parentheses indicate the uncertainties in the last digits. The inset shows $P[I_3 = 0]$ as a function of $p$. (b) The probability distribution of the purification time, $P[\tau]$, near the critical point ($p = 0.119, u = 0.5$) and its log-normal fitting. The inset shows the size $L$ dependence of the median purification time $\tau_{\text{med}}$ near the transition, exhibiting $\tau_{\text{med}}\sim L^{0.95}$ at the critical point $(p , u) = (0.119, 0.5)$. Each $\tau_{\text{med}}$ is extracted from $5,000$ independent evolutions of the circuits initialized in the FOIC with a single ancilla qubit.}
\end{figure}

To precisely investigate the criticality of the MIPT, we perform a finite-size scaling collapse on the probability $P[I_3 = 0]$, where the tripartite mutual information becomes exactly zero, as a function of the rescaled variable $\left(p - p_{c_{1}}\right)L^{1/\nu_{\perp, c_{1}}}$. For this analysis, the double-flag circuits initialized in the FOIC are evolved up to $T=2L$ at a fixed unflagging rate $u > 0$ for various system sizes ranging from $L = 32$ to $1,024$. As shown in Fig.~\ref{fig:MIPT_0_5}, the data for different system sizes successfully collapse onto a single universal curve, yielding the critical point $p_{c_{1}} = 0.119(1)$ and the correlation length exponent $\nu_{\perp, c_{1}}=1.32(3)$. These values are consistent with those reported in conventional MIPTs~\cite{Li2019, Skinner2019, Gullans2020a, Gullans2020b}. We also verified that varying the unflagging rate $u$ merely shifts the critical point $p_{c_{1}}$, leaving the critical exponents of the MIPT unchanged (details not shown).

\begin{figure*}
\includegraphics[width=0.95\textwidth]{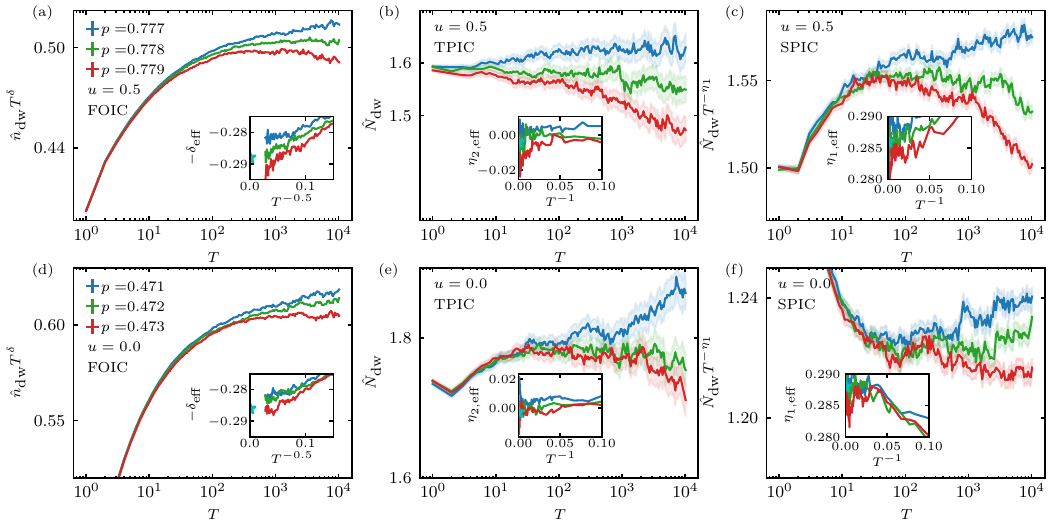}%
\caption{\label{fig:APT_OP} Semilogarithmic plots of order parameters for the absorbing phase transition (APT) at fixed unflagging rates $u = 0.5$ (top row) and $u = 0.0$ (bottom row). The columns show the time evolution of the domain-wall density $\hat{n}_{\text{dw}}$ initialized in the FOIC [(a), (d)]; the average number of domain walls $\hat{N}_{\text{dw}}$ initialized in the TPIC [(b), (e)]; and SPIC [(c), (f)]. Each inset displays the effective exponent of the corresponding observable: $-\delta_{\text{eff}}$ for $\hat{n}_{\text{dw}}$, $\eta_{2,\text{eff}}$ for $\hat{N}_{\text{dw}}$ (TPIC), and $\eta_{1,\text{eff}}$ for $\hat{N}_{\text{dw}}$ (SPIC). At the critical point, the effective exponent approaches its asymptotic value as $T \to \infty$. In contrast, for $p < p_{c_{2}}$ ($p > p_{c_{2}}$), the curve veers upward (downward), implying that the system is in the active (absorbing) phase.}
\end{figure*}

Furthermore, to determine the dynamical exponent $z$ of the MIPT, we investigate the system-size dependence of the purification time $\tau$. This is defined as the time required for an ancilla qubit, initially maximally entangled with a single system qubit, to completely purify (i.e., $\langle S_Q\rangle$ vanishes). Since the probability distribution of $\tau$ near the critical point exhibits a heavy tail closely resembling a log-normal distribution [see Fig.~\ref{fig:MIPT_0_5}(b)], we characterize the scaling behavior using the median purification time $\tau_{\text{med}}$. As illustrated in the inset of Fig.~\ref{fig:MIPT_0_5}(b), we observe a scaling $\tau_{\text{med}}\sim L^{0.95}$, which confirms $z\approx1$ and is also in excellent agreement with conventional MIPTs.

\subsection{\label{sec:APT}Absorbing-state phase transition}

Next, we turn our attention to the APT that occurs at larger measurement rates. To analyze this transition, we first evaluate the estimates of the critical exponents $\delta$, $\eta_2$, and $\eta_1$ corresponding to three initial conditions: the FOIC, TPIC, and SPIC, respectively. The ensemble averages for the observables are computed over $20,000$ independent circuit realizations evolving up to a maximum time $T_{160}=10,240$ to obtain the domain-wall density $\hat{n}_{\text{dw}}$ in the static simulations ($L = 2,048$), and over $100,000$ realizations for the average number of domain walls $\hat{N}_{\text{dw}}$ in the defect simulations. To precisely determine the critical exponents, we extract the LCSE for the static simulations, $\chi_{\hat{n}} = 0.5$, via the LCSE function $\Theta_{\hat{n}}$, while it could not be reliably determined for the defect simulations (details not shown) . This occurs because the initial number of domain walls in the FOIC is approximately a factor of $L$ larger than in the TPIC or SPIC, assuming the same number of realizations. This elevated noise hinders a reliable extraction of the LCSE function $\Theta_{\hat{N}}$ for defect simulations using only $100,000$ realizations.

By evaluating the effective exponents, we obtain $\delta = 0.288(3)$, $\eta_2 = -0.004(5)$, and $\eta_1 = 0.286(5)$ at the critical measurement rate $p_{c_{2}} = 0.778(1)$ for $u = 0.5$ [Figs.~\ref{fig:APT_OP}(a)--(c)]. These values are consistent with the results of previous studies of the DI class~\cite{Hinrichsen2000, Odor2004, Jensen1994, Menyhard2000, Park2020b, Grassberger2026}. Similarly, as shown in Figs.~\ref{fig:APT_OP}(d)--(f), the simulations yield consistent effective exponents $\delta = 0.286(3)$, $\eta_2 = -0.003(4)$, and $\eta_1 = 0.288(5)$ at the critical measurement rate $p_{c_{2}} = 0.472(1)$ for $u = 0.0$, suggesting that the APT belongs to the DI universality class regardless of the unflagging operations.

\begin{figure}
\includegraphics[width=0.48\textwidth]{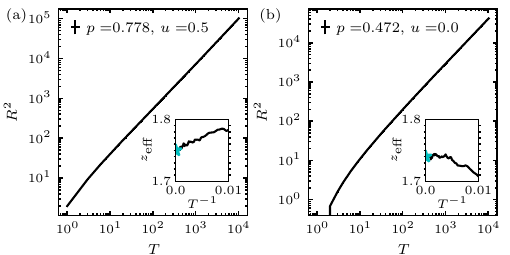}%
\caption{\label{fig:z_APT} Mean squared distance $R^2$ between the two most distant domain walls initialized from the SPIC at two critical points $(p, u)$, (a) $(0.778,0.5)$ and (b) $(0.472,0.0)$. Each inset shows the effective dynamical exponent $z_{\text{eff}}$ as a function of $T^{-1}$. The estimate for $z$ is obtained from the $y$-intercept ($T^{-1} \to 0$) of the critical curve.}
\end{figure}

Having determined the critical point $p_{c_{2}}$ from Fig.~\ref{fig:APT_OP}, we investigate the time evolution of the mean squared distance $R^2$ between the two most distant domain walls to extract the dynamical exponent $z$. By extrapolating the effective exponent $z_{\text{eff}}$ to the infinite-time limit (i.e., $T^{-1}\to0$), as shown in the insets of Fig.~\ref{fig:z_APT}, we determine $z_{c_{2}} = 1.75(1)$ at $p_{c_{2}} = 0.778$ for $u = 0.5$ and $z_{c_{2}} = 1.74(2)$ at $p_{c_{2}} = 0.472$ for $u = 0.0$. These consistent estimates further corroborate that the APT belongs to the DI universality class. Moreover, using the scaling relations between critical exponents, we derive the estimates for the remaining critical exponents: $\nu_{\perp,c_{2}} = (\delta z_{c_{2}})^{-1}=2.0(1)$, $\nu_{\parallel, c_{2}} = z_{c_{2}}\nu_{\perp,c_{2}}=3.5(2)$, and $\beta_{c_{2}} = \delta \nu_{\parallel,c_{2}} = 1.0(1)$.

While the APT robustly belongs to the DI universality class regardless of the unflagging rate $u$, tracking the shift in the critical measurement rate $p_{c_{2}}(u)$ suggests the presence of a nontrivial crossover behavior. This crossover can be described by the following scaling ansatz~\cite{Park2006, Park2007, Park2008, Park2020a}:
\begin{equation}
    \hat{n}_{\text{dw}} (T; p, u) = T^{-\delta}\mathcal{G} \left(T u^{\mu_{\parallel}}, T^{1/\nu_{\parallel,c_{2}}} (p - p_{c_{2}}(0)) \right),
\end{equation}
where $p_{c_{2}}(0)$, $\delta$, and $\nu_{\parallel, c_{2}}$ are the critical parameters obtained above, $\mathcal{G}$ is a universal scaling function, and $\mu_{\parallel} = \nu_{\parallel, c_{2}} / \phi$ with $\phi$ being the crossover exponent. For sufficiently small $u$, the phase boundary is governed by the crossover exponent, implying that the shift in the critical measurement rate scales as
\begin{equation}
    p_{c_{2}}(u) - p_{c_{2}}(0) \sim u^{1/\phi}.
\end{equation}
Furthermore, precisely at $p = p_{c_{2}}(0)$, the second argument of the scaling function vanishes, yielding
\begin{equation}
\label{eq:SA_CO}
    \hat{n}_{\text{dw}} (T; p_{c_{2}}(0), u) T^{\delta} = \mathcal{G} \left(T u^{\mu_{\parallel}}, 0 \right).
\end{equation}

\begin{figure}
\includegraphics[width=0.48\textwidth]{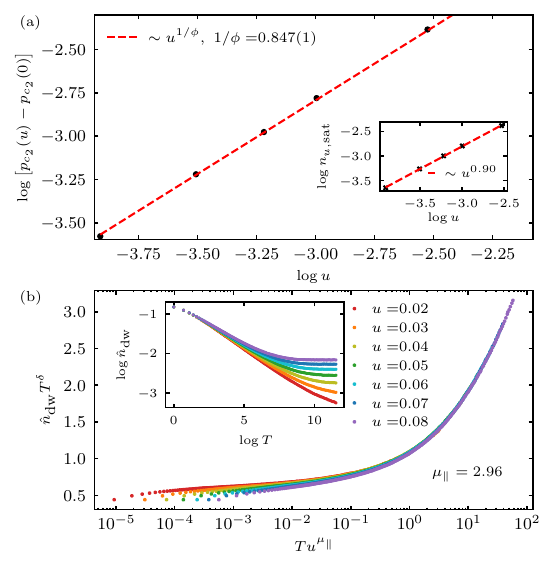}%
\caption{\label{fig:NTCO} Nontrivial crossover behavior at small unflagging rates $u$. (a) Log-log plot of the shift in the critical measurement rate, $p_{c_{2}}(u)-p_{c_{2}}(0)$, as a function of $u$. The red dashed line indicates a power-law fit yielding the crossover exponent $1/\phi = 0.847(1)$. The inset shows the saturated proportion of unflagged domain-wall flags $n_{u,\text{sat}}$ at critical points as a function of $u$ with a power-law fitted red line, yielding $n_{u,\text{sat}} \sim u^{0.9}$. (b) Semilogarithmic plot showing the scaling collapse of $\hat{n}_{\text{dw}} T^{\delta}$ versus the rescaled time $T u^{\mu_{\parallel}}$ with $\mu_{\parallel}=2.96$, evaluated for $u$ ranging from $0.02$ to $0.08$ at a fixed measurement rate $p=0.472=p_{c_{2}}(u = 0)$. The inset represents the time evolution of $\hat{n}_{\text{dw}}$ up to a maximum time $T_{210}=102,400$ for the corresponding $u$ values.}
\end{figure}

As depicted in Fig.~\ref{fig:NTCO}(a), the scaling of $p_{c_{2}}$ with respect to $u$ is confirmed via a power-law fit, which yields $1/\phi = 0.847(1)$, or equivalently, $\phi = 1.181(2)$. To independently verify this, we perform a scaling collapse based on Eq.~\eqref{eq:SA_CO}, as shown in Fig.~\ref{fig:NTCO}(b). For this analysis, circuits of size $L=2,048$ initialized in the FOIC are evolved up to $T_{210} = 102,400$ at a fixed measurement rate $p = 0.472 = p_{c_{2}}(0)$. Extracting $\hat{n}_{\text{dw}}$ averaged over $10,000$ independent realizations, the data collapse onto a single curve against the rescaled time $T u^{\mu_{\parallel}}$ with $\mu_{\parallel}=2.96$. This estimate is highly consistent with the previously obtained exponents, satisfying the scaling relation $\mu_{\parallel} \approx 3.5 \times 0.847 \approx 2.96$.

\begin{figure}
\includegraphics[width=0.48\textwidth]{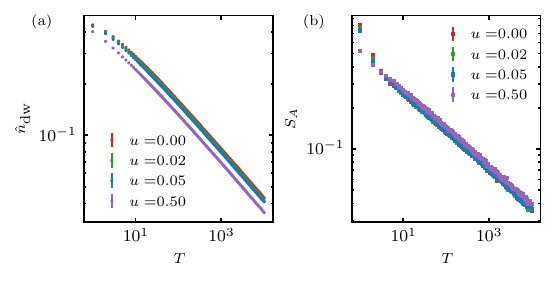}%
\caption{\label{fig:n_dw_EE_different_u} Identical critical scaling of domain-wall density and entanglement entropy across different critical points initialized from the FOIC. (a) Log-log plot of the domain-wall density $\hat{n}_{\text{dw}}$ and (b) bipartite entanglement entropy $S_A$ as a function of time $T$. Each line corresponds to the critical point at different $u$ (red for $p=0.472,~u=0$, green for $p=0.500,~u=0.02$, blue for $p=0.534,~u=0.05$, and purple for $p=0.778,~u=0.5$, respectively), averaged over $40,000$ different realizations.}
\end{figure}

It is noteworthy that, since all APTs in this circuit belong to the DI class regardless of the unflagging operations, these observations suggest a nontrivial crossover occurring within the identical universality class, reminiscent of the behavior reported in Ref.~\cite{Park2007}. In that study, a nontrivial crossover within the DP class was shown to be driven by an abrupt change in the auxiliary field density, resulting in the reduction of the number of absorbing states—specifically, between a model with a single absorbing state and one with infinitely many—even though both exhibit DP critical scaling. This implies that an analogous mechanism might underlie the crossover observed in our DI system.

Further insight is gained by examining the evolution of the domain-wall flags $\{g_i\}$ modified by the unflagging operations evaluated from the FOIC. Specifically, we define the proportion of additional domain-wall flags generated by unflagging as $n_u \equiv 1 - N_f / N_g$ (averaged over $10,000$ realizations, within which no trajectory reaches the absorbing state $N_g = 0$), where $N_f$ denotes the number of mismatched adjacent pairs inferred from the measurement flags $\{f_i\}$, and $N_g$ represents the total number of $0$'s in $\{g_i\}$. At the critical points, $n_u$ saturates in the long-time limit to a steady value scaling as $n_{u,\text{sat}} \sim u^{0.9}$ [see the inset of Fig.~\ref{fig:NTCO}(a)], suggesting that this quantity may play an analogous role to the emergence of a finite auxiliary field density responsible for driving the crossover, given that $n_u = 0$ when $u=0$. Moreover, quantum entanglement is unlikely to play a significant role in this crossover behavior because the critical line of the APT is located deep in the area-law phase. Specifically, when initialized in the FOIC, the bipartite entanglement entropy $S_A$ at the critical points decays as $S_A \sim T^{-\delta}$ regardless of the unflagging rate $u$, as shown in Fig.~\ref{fig:n_dw_EE_different_u}(b). This scaling mirrors the decay of the domain-wall density $\hat{n}_{\text{dw}} \sim T^{-\delta}$ [Fig.~\ref{fig:n_dw_EE_different_u}(a)], indicating that the entanglement dynamics at the APT criticality is primarily governed by the diminishing population of domain walls. Whether the observed scaling $1/\phi = 0.847(1)$ (alongside $n_{u,\text{sat}} \sim u^{0.9}$) reflects an asymptotic crossover or an effective exponent governed by strong finite-time corrections to scaling in the small-$u$ regime, as well as the precise microscopic origin of this behavior, poses an intriguing direction for future research.

\section{\label{sec:discussion_conclusion}Discussion \& Conclusion}
In summary, we have investigated two types of nonequilibrium phase transitions in double-flag circuits, which steer the system toward $\mathbb{Z}_2$-symmetric states. In the absence of unflagging operations ($u = 0$), a sufficiently large measurement rate $p$ drives the system directly into the absorbing state without undergoing an entanglement phase transition. However, when unflagging operations are present—facilitating entanglement generation by preventing the system's domain-wall density $\hat{n}_{\text{dw}}$ from rapidly decaying—the circuit exhibits an MIPT and an APT sequentially as $p$ increases. By evaluating the critical exponents, we confirmed that the universality classes of these transitions remain robust against variations in the unflagging rate for $u > 0$. Specifically, the critical exponents of the APT extracted from three distinct initial conditions are highly consistent with previously reported values. This confirms that the APT belongs to the DI universality class, highlighting a strong classical-quantum correspondence between branching-annihilating random walks with an even number of offspring and these monitored Clifford circuits with feedback.

Analogous to the quantum order parameters, one could define classical order parameters for the APT using the domain-wall flags: $n_{\text{dw}}\equiv 1 -\frac{1}{L}\sum_{i=1}^{L}g_i$, $N_{\text{dw}}\equiv L-1-\sum_{i=1}^{L-1} g_i$. Intriguingly, despite the inherent discrepancies between the classical and quantum observables, we find that $n_{\text{dw}}$ and $\hat{n}_{\text{dw}}$ yield consistent results for examining the criticality of the DI class. This suggests that the classical order parameters provide reliable proxies for characterizing the critical behavior of the quantum system, effectively bypassing the need for direct quantum state measurements as the system evolves.

In conclusion, the double-flag circuit is highly adaptable. For instance, to directly examine the order parameter exponent $\beta_{c_{2}}$, one could design circuits to evaluate the steady-state domain-wall density $\hat{n}_{\text{s}}$ near the critical measurement rate $p_{c_{2}}$. Moreover, by intentionally applying feedback-control operations that break the $\mathbb{Z}_2$ symmetry (e.g., a layer of CNOT gates) with a probability $q$, one could engineer a quantum circuit that exhibits a crossover from the DI transition to a DP transition as $q$ is varied~\cite{Park2008}. Alternatively, introducing static disorder into the measurement rate could allow one to verify whether $\nu_{\perp, c_{2}} = 2$ exactly~\cite{Zabalo2023} by analyzing the dynamical exponent $z$ as a function of the disorder strength. However, exploring these avenues is computationally demanding, as it requires an extensive number of realizations for circuits with significantly larger $L$ and $T_{\text{max}}$. Therefore, we leave these compelling directions for future work.

\begin{acknowledgments}
This work was supported by the National Research Foundation (NRF) of Korea grant funded by the Korea government (MSIT) (No. RS-2024-00392445) and the Regional Innovation System \& Education (RISE) program through the Gwangju RISE Center, funded by the Ministry of Education (MOE) and the Gwangju Metropolitan City, Republic of Korea (2026-RISE-05-001).
\end{acknowledgments}

\bibliography{paper}

\end{document}